\documentstyle[preprint,aps]{revtex}
\begin{document}
\draft
\title{Efficient Mixing at low Reynolds numbers using polymer additives}
\author{Alexander Groisman$^1$ and Victor Steinberg}
\address{
Department of Physics of Complex Systems\\
The Weizmann Institute of Science, 76100 Rehovot, Israel,\\
$^1$ Present address: Department of Applied Physics, Caltech, Pasadena, california 91125, USA.}
\date{\today}
\maketitle

\narrowtext

{\bf Mixing in fluids is 
a rapidly developing field of fluid mechanics \cite{Sreen,Shr,War}, being an important industrial
and environmental problem. The mixing of liquids at low Reynolds numbers is usually quite weak 
in simple flows, and it
requires special devices to be efficient. Recently, the problem of 
mixing was solved analytically for a simple case of random flow, known as the
Batchelor regime \cite{Bat,Kraich,Fal,Sig,Fouxon}. Here we demonstrate experimentally that very
viscous liquids at low Reynolds number, $Re$.  Here we show
that very viscous liquids containing a small amount of high molecular weight polymers can be 
mixed quite efficiently at very low Reynolds numbers, for a simple flow in
a curved channel. A polymer 
concentration of only 0.001\% suffices.
The presence of the polymers leads to an elastic instability \cite{LMS} and to 
irregular flow \cite{Ours}, with velocity spectra corresponding to the Batchelor regime 
\cite{Bat,Kraich,Fal,Sig,Fouxon}. Our detailed observations of the mixing in this regime
enable us to confirm sevearl important theoretical predictions: the probability distributions 
of the  concentration exhibit exponential tails \cite{Fal,Fouxon}, moments of the 
distribution decay exponentially along the flow \cite{Fouxon}, and the spatial correlation 
function of concentration decays  logarithmically.}

Solutions of flexible high molecular weight polymers are viscoelastic fluids \cite{Bird}. 
There are elastic stresses that appear in the polymer solutions in a flow, and that grow
non-linearly with the flow rate \cite{Bird}. This can lead to many special flow effects, 
including purely elastic transitions \cite{LMS,McK,Joo} that qualitatively change character 
of the flow at vanishingly small $Re$. As a result of such
transitions secondary vortical flows appear in different systems, where the primary motion is
a curvilinear shear flow. Onset of those secondary flows depends on the Weissenberg
number, $Wi=\lambda\dot{\gamma}$, where $\lambda$ is the polymer relaxation time and 
$\dot{\gamma}$ is the shear rate. It plays a role analogous to that of $Re$ in
competition between non-linearity and dissipation. As it has been reported recently, an 
elastic flow transition can result in a special kind of turbulent motion, elastic turbulence 
\cite{Ours}, which arises at arbitrary small $Re$. In our experiments we find that
irregular flow excited by the elastic stresses in the polymer solutions can lead 
to quite efficient mixing at very low $Re$.

Our mixing experiments were carried out in a curvilinear channel, Fig.1a, 
at a room temperature, $22.5 \pm 0.5$ $^{\circ}$C. The total rate of the fluid
discharge, $Q$, was always kept constant, so that the average time of mixing was 
proportional to the position, $N$, along the channel, Fig.1a.
We used a solution of 65\% saccharose and 1\% NaCl in water, viscosity 
$\eta_s=0.153$ Pa s, density $\rho=1.32$ g/cm$^3$, as a solvent for the polymer. 
We added polyacrylamide, $M_w=18,000,000$, at a concentration of 
80 p.p.m. by weight. One of the solutions entering the channel, Fig.1a, also contained
$c_0=2$ p.p.m. of a fluorescent dye. The
solution viscosity was $\eta=0.198$ Pa$\cdot$s at a shear rate $\dot{\gamma}=4$ s$^{-1}$.
The relaxation time, $\lambda$, estimated from the phase shift between the stress and
the shear rate in oscillatory tests, was 1.4 s. An estimate for the diffusion
coefficient of the dye is given by that for the saccharose molecules, which is about 
$D=8.5\cdot10^{-7}$ cm$^2$/s. The characteristic shear rate and the Weissenberg number
in the flow are estimated as $\dot{\gamma}=(2Q/d^2)/(d/2)=4Q/d^3$ and 
$Wi=\lambda(4Q/d^3)$, respectively.

The Reynolds number, $Re=2Q\rho/(d\eta)$, was
always quite small. It only reached 0.6 for highest $Q$ that we explored. 
Therefore, flow of the pure solvent remained quite laminar and no mixing occurred, Fig.1b. 
The boundary separating the liquid with and without the dye was smooth and parallel 
to the direction of the flow, and it only became smeared due to molecular diffusion
as the liquids advanced downstream. 
Behaviour of the polymer solution was qualitatively different from that of the solvent. 
The flow only was laminar and stationary up to a value of $Q$ corresponding
to $Wi_c=3.2$ (and $Re=0.06$), at which an elastic instability 
occurred. This instability lead to irregular flow and
mixing of the liquids, Fig.1c. The experiments were carried out at $Q$ about twice 
above the flow instability onset, $Wi=6.7$, at which homogeneity 
of the mixture at the exit of the channel was the highest.

Power spectra of fluctuations of velocity in the center of the channel at $Wi=6.7$
are shown in Fig.2. The spectra of both longitudinal and transversal velocity components 
do not exhibit any distinct peaks and have broad regions of a power law decay, that is 
typical for turbulent flow. They resemble the power spectra found at similar $Wi$ in 
flow of the same polymer solution between two parallel plates in the regime of 
elastic turbulence \cite{Ours}. So, we believe the origin of the irregular flow is the
same here as in Ref. \cite{Ours}.

According to the Taylor hypothesis fluctuations of the flow velocity
in time are mainly due to fluctuations in space advected by the mean flow \cite{Shr,War}. 
So, the spectra in Fig.2 imply that the power of the velocity fluctuations scales with 
the $k$-number in space as $P\sim k^{-3.3}$. Fluctuations of the velocity gradients 
should scale then as $k^{-1.3}$, so that the flow becomes increasingly homogeneous at 
small scales, and mixing is mainly due to the largest eddies having the size of the whole flow 
system \cite{Fal}. Such flow conditions correspond to the Batchelor regime of mixing \cite{Bat,Kraich}.
It is one of the two simple cases of random flow, where the problem of mixing has been solved 
analytically \cite{Shr,Kraich,Fal,Sig,Fouxon}. Investigation of mixing in the irregular flow
in the channel gave us an opportunity to study mixing in the Batchelor regime in detail and to test 
a few key the theoretical predictions for the first time.

Mixing of the polymer solution was 
a random process. So, it was appropriate to characterize it statistically, 
by probability distribution function (PDF) to find different concentrations,
$c$, of the dye in a point, and by values of the moments, $M_i$, of the distribution. 
An $i$-th moment is defined as an average,
$<|c-\bar{c}|^i>/\bar{c}^i$, where $\bar{c}$ is the average concentration of the dye, which 
in our case is equal to $c_0/2$. Small values of $M_i$ mean high homogeneity and 
good mixing of the liquids. $M_i$ are all unity at the channel entrance
and they become zero for perfectly mixed liquids. 

PDF of the concentrations at $N=8$ and $N=24$ (with $M_1=0.25$ and 0.72, respectively) are 
shown in Fig. 3a-b. Dependencies of $M_1$ and 
$M_2$ on the position, $N$, along the channel are presented in Fig.4.
Advanced stages of mixing corresponding to $N>30$ were studied,
when the liquids were premixed before they entered the channel, Fig.4. Representative 
PDF in the flow with preliminary mixing taken at the positions, where $M_1=0.082$ and $0.030$, 
are shown in Fig.3c-d, respectively.

Representative spatial autocorrelation functions for the dye concentration are shown in Fig.5.
At the beginning the concentration
distribution is strongly influenced by the initial conditions. There are extended uniform 
regions with maximal and zero dye concentration, so, the concentration remains highly correlated 
over rather large distances,
Fig.5a, and PDF has maxima near $c_0$ and zero, Fig.3a. As the liquid advances downstream, 
it becomes increasingly homogeneous. Mixing patterns exhibit a lot of fine scale
structures of different brightnesses, the correlation distances become shorter, Fig.5b, 
and PDF of the dye concentration become narrower, Fig.3b. PDF in Fig.3b
has a single peak at $\bar{c}$ and long tails that decay exponentially and touch the limits
of the concentration, zero and $c_0$. 

Further downstream mixing patterns have characteristic features at similar spatial scales, 
Fig.5c-e, but are much more faded. The PDF in Fig.3c is much narrower than in Fig.3b and has quite clear 
exponential tails, in agreement with the theoretical predictions \cite{Fal,Sig}, that implies 
significant intermittency in mixing \cite{Shr}. The distribution is
well confined in a region far from the limits of zero and $c_0$. So, the dependence 
on the initial conditions should be quite minor by that point. At the last point 
($N=54$, $M_1=0.030$) the non-homogeneity in the concentration is hardly seen, and  
the PDF, Fig.3d, is very narrow. One can see that at $N$ of about 30 the autocorrelation 
function reaches some asymptotic form, which does not
change as the liquid moves further downstream. This may correspond to a situation, when the 
small scale structures are smeared by the molecular diffusion at the same rate as they are
created by the flow. The autocorrelation functions, Fig.5c-e, decay logarithmically at the 
distances above the diffusion length, that is in agreement with the theoretical predictions 
for the Batchelor regime \cite{Bat,Kraich,Fal}. Spatial structure of concentration fluctuations
in the Batchelor regime has been experimentally studied before in different types of flows 
\cite{War,Dim,Gollub}. However, agreement between experiment and theory remained rather controversial
(see Ref.2, 14 and 15 for discussion).

The dependencies of $M_1$ and $M_2$ on $N$ in Fig.4 have inflections at $N$ between 
20 and 30. From the above discussion we can suggest that they are due to
transition to an asymptotic regime, where dependence on the initial conditions is lost. 
The correlation functions at different positions become identical, and PDF 
of concentrations can be superimposed rather well for $\Delta c$ of about $3M_1\bar{c}$ around
$\bar{c}$ by appropriate stretching of the axes. There is no self-similarity in the shapes 
of PDF at larger $\Delta c/(M_1\bar{c})$, however, and the shape permanently changes with $N$. 
One can learn from Fig.4 
that both $M_1$ and $M_2$ decay exponentially above $N=30$, the rate of the decay being two 
times higher for $M_2$ than for $M_1$. The higher order moments were found to decay 
exponentially, $M_i\sim exp(-\gamma_i N)$, as well, that is quite in agreement
with the theoretical predictions \cite{Fouxon}.  We note here that $M_2$, which is the variance
of fluctuations of the dye concentration, is often considered as an
analogue of energy of turbulent motion \cite{Shr}. So its logarithmic derivative, 
${d ln M_2 \over dt}=-\gamma_2/7.8$ s$^{-1}$ (see Fig.4), is the analogue of the energy 
dissipation rate. The theory also predicts linear growth of the coefficients, $\gamma_i$,
with $i$ at small $i$ and saturation of the growth, when $i$ becomes large \cite{Fouxon}. 
This form of dependence of $\gamma_i$ on $i$ is quite confirmed by our experimental 
results, inset in Fig.4. The deviation of the dependence of 
$\gamma_i$ on $i$ from a straight line at large $i$ is a quantitative evidence for the 
lack of self-similarity in the mixing. One can see that all our experimental results agree 
very well with all the theoretical predictions for the Batchelor regime of mixing, we tested
\cite{Bat,Kraich,Fal,Sig,Fouxon}.

As it is suggested by the decay of $M_1$ in Fig.4, the polymer solutions are mixed at 
a characteristic distance of $\Delta N\simeq 15$. It corresponds to the total path of
about $140d$ and the average flow time of about 120 seconds, which is three orders
of magnitude smaller than the diffusion time, $d^2/D$. Using a more concentrated sugar 
syrup as a solvent, we prepared a polymer solution with $D$ about 30\% smaller and with 
viscosity and relaxation time about two times larger than those of the original solution. 
We studied distribution of $c$ at $N=29$ in a flow of this solution at the same $Wi$ of 6.7 
as before. Here the characteristic flow velocities were twice lower, $Re$ was about four times 
lower, and the ratio between the flow time and the diffusion time, $d^2/D$, was about 1.5 
higher than in the original flow. The measured values of $M_1$ and $M_2$ were the same as in Fig.4. 
So, the inertial forces and the molecular diffusion did not have any apparent influence on the
mixing efficiency. 
Dependence of the 
efficiency of mixing at the optimal flow conditions on concentration of the polymers was 
surprisingly weak (although $Wi_c$ grew fast, when the polymer concentration was decreasing). 
So, for a solution with the polymer concentration of 10 p.p.m. ($\eta/\eta_s=1.03$), $M_1$
of as low as 0.22 was reached at $N=29$ (and at $Re=0.065$). The mixing was observed down 
to the polymer concentration of 7 p.p.m. Thus, very viscous liquids can be efficiently mixed 
in curvilinear channels at very low flow rates with the aid of polymer additives 
at very low concentrations. This method of mixing, we believe, can find some 
industrial and laboratorial applications.

\begin{figure}
\caption{Experimental set up and two snapshots of the flow. 
{\bf a)} Experimental set-up. A channel of a depth $d=3$ mm is machined in a 
transparent bar of perspex and sealed from above by a transparent window. The channel is 
square in the cross-section, and consists of a
sequence of 60 smoothly connected half-rings with the inner and outer radii $R_1=3$ mm
and $R_2=6$ mm, respectively. The flow is always observed near the middle of a half-ring on 
the right-hand side of the bar, when viewed streams. 
The half-rings on the right-hand side are numbered starting from the channel entrance,
and the number, $N$, of the half-ring is a natural linear coordinate along the channel. Two liquids are fed 
into the channel by two syringe pumps at equal discharge rates through two separate tubes.
The two working liquids are always identical with the only difference of a small amount 
of a fluorescent dye (fluorescein) added to one of them. The channel is illuminated from 
a side by an Argon-Ion laser beam converted by two cylindrical lenses to a broad
sheet of light, with a thickness of about 40 $\mu$m in the region of observation.
The fluorescent light emitted by the liquid in the perpendicular direction is projected onto
a CCD camera and digitized by a 8-bit, 512$\times$512 frame grabber. Concentration 
of the dye is evaluated from the intensity of the light, which was found to be proportional 
to the concentration.
{\bf b-c} Snapshots of the flow at $N=29$. Bright regions correspond to 
high concentration of the fluorescent dye. Curved boundaries of the channel are seen 
on the left and on the right. {\bf b)} Pure solvent at $Re=0.16$. {\bf c)} Polymer solution
at the same flow rate, corresponding to $Wi=6.7$.}
\end{figure}

\begin{figure}
\caption{Power, $P$, of fluctuations of velocity in the middle of the channel at $N=12$
as a function of frequency, $f$. The flow velocity was measured by a 
laser Doppler anemometer, when the region of 
the laser beam crossing was made very small, $15\times 15\times 40$ $\mu$m, to decrease the
gradient noise. The spectra of the velocity components for the polymer solution along 
the mean flow, curve $a$, across the mean flow, curve $b$, and for the pure solvent across 
the mean flow, curve $c$, are shown. The mean velocity for the polymer solution
was $\bar{V}=6.6$ mm/s. The RMS of fluctuations, $V_{rms}$, was $0.09\bar{V}$ and 
$0.04\bar{V}$ in the longitudinal and transversal directions, respectively.}
\end{figure}

\begin{figure}
\caption{Plots of PDF of concentration of the fluorescent dye at different positions. 
Brightness profile was taken 12.5 times per 
second along a single line across the channel near the middle of a half-ring (a horizontal 
line in the middle in Fig.1b-c). Each plot represents statistics over about $10^7$ points, 
corresponding to about 25,000 different moments of time, and a total liquid discharge 
of $2\cdot10^3d^3$. The regions near the walls of the channel
with the widths of $0.1d$ were excluded from the statistics. $a-b$ -
PDF at $N=8$ and $N=24$; $c-d$ - with preliminary mixing, at the positions 
corresponding to $N=39$ and $N=54$, respectively.}
\end{figure}

\begin{figure}
\caption{Dependence of $M_1$ and $M_2$, circles and squares, respectively, 
on the position, $N$, along the channel. The average flow time is connected to $N$ 
as $\bar{t}=N\cdot 7.8$ s. In order to observe stages of mixing corresponding to $N>30$,
we carried out a series of experiments, where the liquids were premixed before they entered 
the channel. A shorter channel of the same shape was designed for this purpose and put before the entrance 
to the original one. As a result of this preliminary mixing, PDF of the dye concentrations
at $N=2$ was almost identical to PDF at $N=27$ without the premixing. 
Filled symbols represent $M_1$ and $M_2$ without preliminary
mixing. Empty symbols are for the flow
with preliminary mixing, when a value of 25 is added to the actual position, 
in order to match the entrance conditions. . One can see that these curves are indeed continuations of the 
dependencies obtained for $M_1$ and $M_2$ in the channel without the premixing.}
\end{figure}

\begin{figure}
\caption{Correlation coefficients for the concentration as functions of the distance, 
$\Delta x$, across the channel (semilogarithmic coordinates). The curves $a-d$ correspond 
to the same conditions as PDF in Fig.5a-d, and they were calculated using the same data arrays. 
The curve $e$ is for $N=29$. The correlation functions coincide for $N\geq 29$. At
small $\Delta x$ they have parabolic scaling with a characteristic length 
$x_0=0.017d\approx 50$ $\mu$m. It is, probably, defined by the thickness of the
illuminating light sheet, about 40 $\mu$m, and by the molecular diffusion scale, 
$(D/(V_{rms}/d))^{1/2}\simeq 25$ $\mu$m.  At larger $\Delta x$ the scaling is logarithmic.}
\end{figure}


{\bf Acknowledgements.}

We gratefully acknowledge G. Falkovich for theoretical guidance and numerous 
discussions. The work was partially supported by the Minerva Center for Nonlinear 
Physics of Complex Systems, by a Research Grant from the Henry Gutwirth Fund and
by Israel Science Foundation grant.

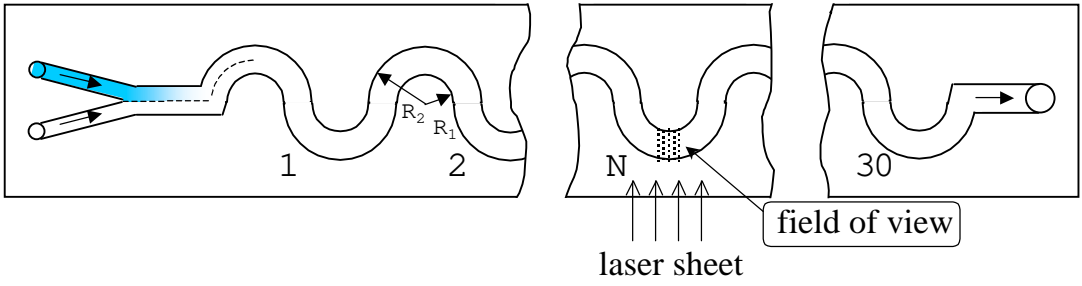
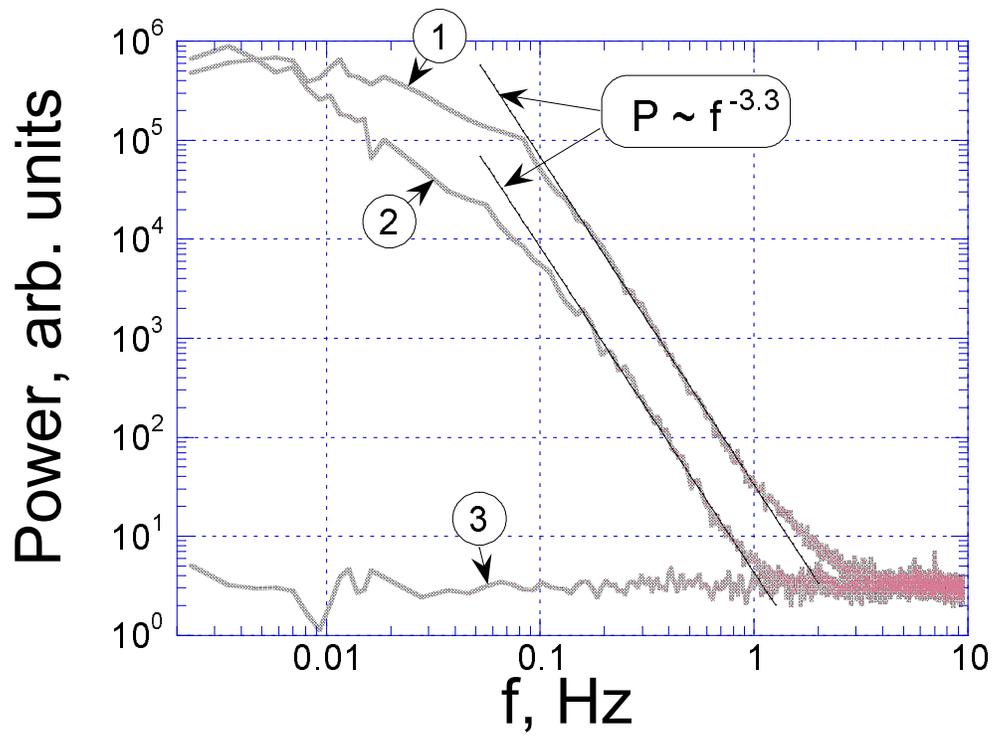
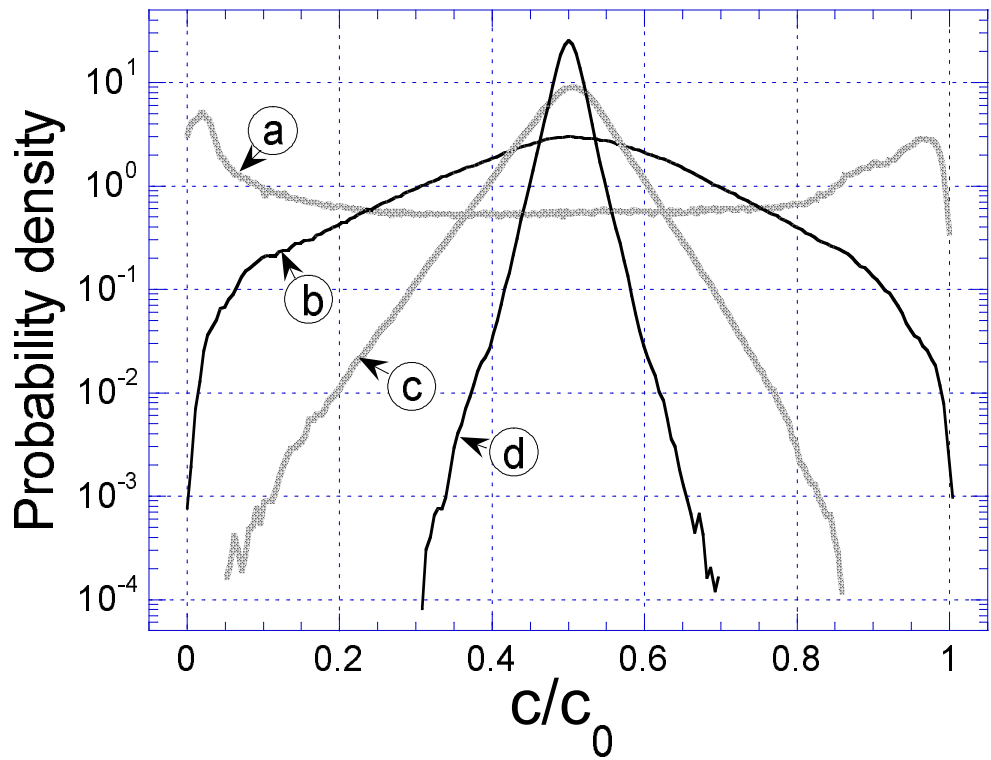
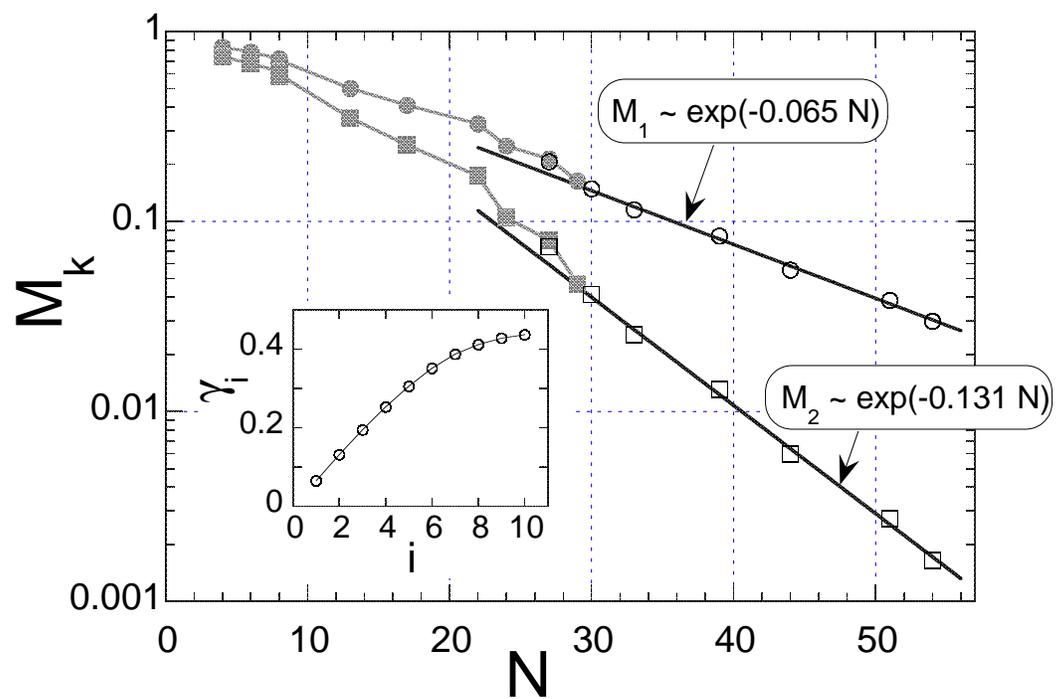
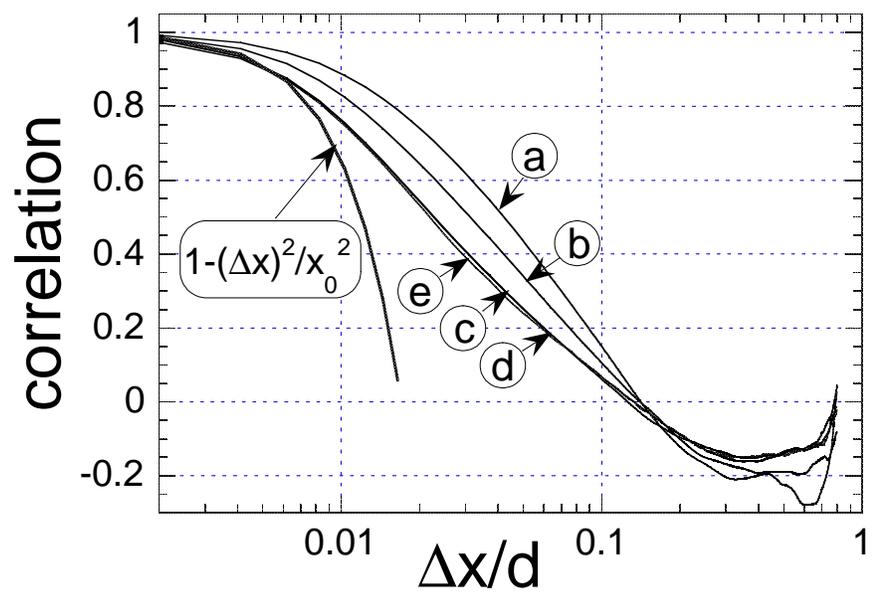
\end{document}